\documentclass[epj]{svjour}
\usepackage{graphics}
\begin{document}
\title{Multiplicities and particle production at LEP\thanks
{Talk 
given at the International Europhysics Conference on High Energy Physics, 
EPS-HEP2003, 17-23 July 2003,  Aachen, Germany.
}
}
\author{Edward K. G. Sarkisyan\inst{1,}\inst{2}%
}                     
%
%
\institute{EP Division, CERN, CH-1211 Geneva 23, Switzerland \and 
Department of Physics and Astronomy, The University of Manchester,
Manchester, M13 9PL, UK}
\date{Received: date / Revised version: date}
%
\abstract{Recent results on 
 hadron multiplicities in heavy and light quark
fragmentation  above 
the Z$^0$ peak (OPAL), and 
multiplicity distribution analysis  (L3) and inclusive $f_1$ production 
(DELPHI) in hadronic Z$^0$ decays are presented.
\PACS{
      {12.38.Qk}{}
\and
      {13.66.Bc}{}
     } 
} 
%
\maketitle
\section{Hadron multiplicities in heavy and light quark
fragmentation} \label{flavour}

\begin{figure}[t]
\resizebox{0.47\textwidth}{0.62\hsize}
{
  \includegraphics{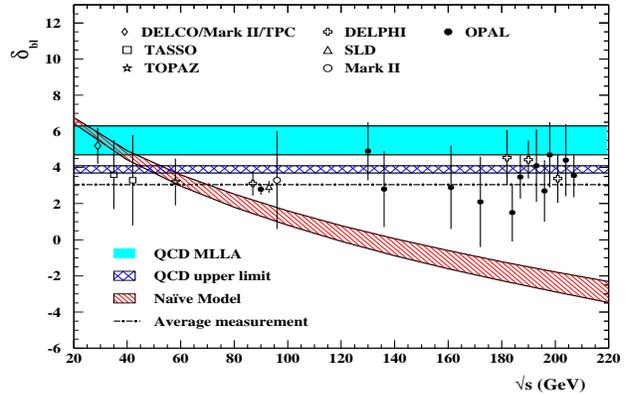}
}
\caption{
The difference in the mean charged multiplicities,
$\delta_{\rm bl}$,
between heavy and light quark pairs as a function of centre-of-mass
energy. The dashed-dotted line is the combined result from 
all
measurements. The following predictions are shown: 
the MLLA prediction \cite{qcdh1} (does not include higher-order 
corrections), the QCD upper limits as in  \cite{qcdh2}, and the naive 
model calculations \cite{naive}. See  \cite{opalbl} for 
more
details.}
\label{fig:opalbl}
\end{figure}

OPAL presents the recent measurements of
charged hadron multiplicities in heavy
and
light quark initiated events  from the full statistics collected at 
LEP1.5 and LEP2
\cite{opalbl}.
The study of the quark content in multiparticle
production provides
one of the basic tests of QCD.
The results from LEP are of a special interest since they
cover a wide centre-of-mass energy region and can be directly compared
with
QCD which is mostly predictable at asymptotic energies \cite{qcdrev}.

In \cite{opalbl}, OPAL performs a study of the fragmentation of
heavy b-quark and light quarks (l = u, d, s). The
measurements of
the difference in charge particle multiplicities,
$\delta_{\rm bl} = \langle n_{{\rm b}{\bar {\rm b}}}\rangle -
                   \langle n_{{\rm l}{\bar {\rm l}}}\rangle$,
 for ${\rm b}{\bar {\rm
b}}$ and ${\rm l}{\bar {\rm
l}}$ events in e$^+$e$^-$ annihilation at the
centre-of-mass energies above the Z$^0$ peak
are carried out. The findings are compared to the theoretical predictions
of
QCD \cite{qcdh1,qcdh2,qcdh3} and to a more phenomenological (the so-called 
na\"\i ve) 
model \cite{naive} 
(for a review see \cite{qcdrev}). The QCD calculations predict 
energy independent
behaviour of the multiplicity difference $\delta_{\rm bl}$, while in the
na\"\i ve model one expects the decrease with increasing energy. The
latter
is connected with the assumption that the hadron multiplicity accompanying 
the heavy hadrons in ${\rm b}{\bar {\rm b}}$ events is the
same as the multiplicity in ${\rm l}{\bar {\rm l}}$ events at the energy 
left to the system once the heavy quarks have fragmented.
The lower energy measurements could not discriminate between the two 
approaches, see Fig. \ref{fig:opalbl}.

The difference between the heavy and light
mean quark-pair multiplicities obtained 
by OPAL
in the
energy
range  up to 206
GeV is 
shown in 
Fig. \ref{fig:opalbl} along with all previously published  results and 
compared to the QCD predictions. OPAL obtains the luminosity
weighted up to 195 GeV 
$\delta_{\rm 
bl}$ average value
 $\delta_{\rm bl}=3.44\pm0.40({\rm stat})\pm0.89({\rm syst})$ 
GeV \cite{opalbl}. This result, which differs 
numerically 
(due to
some
differences in the data processing procedure)
from that  from DELPHI, $\delta_{\rm bl}=4.26\pm0.51({\rm 
stat})\pm0.46({\rm syst})$   
\cite{delbl}, 
  leads to the
conclusion
on the energy independence of $\delta_{\rm bl}$.
This finding
favours the QCD
prediction  while it is inconsistent with the flavour independent 
na\"\i ve
model 
what now is  confirmed by LEP with high 
accuracy.

\section{$H_q$-moment analysis of  the multiplicity distribution in 
hadronic Z decays}
\label{sec:l3}

L3 reports on 
the charged-particle multiplicity study in terms of 
$H_q$ 
moments \cite{l3h}. The $H_q$ moments \cite{hq1} 
are constructed 
as a ratio of 
cumulants to 
factorial moments and look to be more convenient to be 
studied since
they
do not increase so rapidly with  rank $q$ as 
the cumulants or factorial moments do 
\cite{correv}. 
Meanwhile, 
the  
$H_q$s 
 exhibit all the qualitative features of the cumulants, 
particularly their property to extract genuine $q$-particle correlations. 
For a review see 
\cite{multrev}.

The moments $H_q$ appear as the solution of the QCD 
equations for the generating function. Their  $q$-dependence  is 
quite sensitive to the 
approximation used: $H_q =1/q^2$
in the double-log approximation (DLA), they have  
a minimum at 
$q_{\rm min}\approx 5$ in 
the modified leading-log approximation (MLLA or next-to-leading order 
NLO), and   
oscillate around zero for higher NLO terms, e.g. NNLO. The 
minimum and the  
oscillations are observed experimentally in different type of collisions, 
while there is still no clear understanding of the physical origin of the 
oscillations.  Those could appear e.g. due to energy-momentum conservation 
(which is incorporated 
in Monte Carlo models 
and in NNLO), the flavour content, the restrictions of finite energy 
(maximum multiplicity cutoff) etc.

\begin{figure}[t]
\resizebox{0.47\textwidth}{0.62\hsize}{%
  \includegraphics{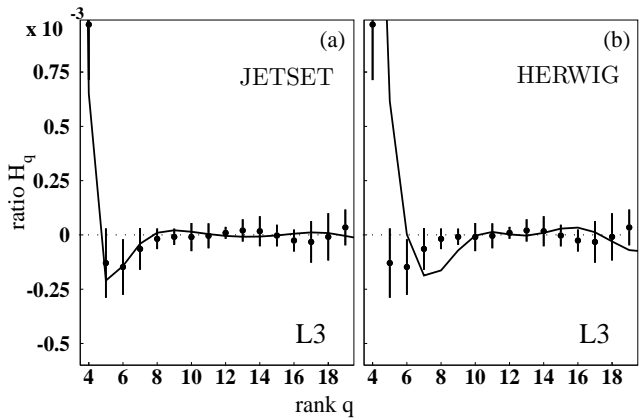}
}
\caption{The $H_q$ moments of the charged-particle multiplicity 
distributions of L3  compared to the predictions of (a) {\sc 
Jetset}  
and (b) 
{\sc Herwig} 
Monte Carlo 
models, as reported in \cite{l3h}.
}%
\label{fig:l3h}      
\end{figure}

 To note is that local parton-hadron duality hypothesis (LPHD) \cite{lphd} 
assumes
that hadronic spectra are proportional to the partonic ones the theory 
(OCD) deals with (for a review see \cite{qcdrev}). 
If this is valid, the same behaviour may be expected for the 
experimentally observed multiplicity distributions as for the parton one. 

Fig. \ref{fig:l3h} shows the $H_q$ moments measured by L3 from the 1.5M Z 
decays sample compared to the expectation of the Monte Carlo 
models,  
{\sc 
Jetset} with string fragmentation and {\sc Herwig} of cluster 
fragmentation. 
The 
moments have a minimum at $q=5$, while the oscillations seem to be  
statistically
insignificant.  
This agrees qualitatively with MLLA and NNLO  but 
does not confirm 
the oscillations predicted. 
{\sc Jetset} agrees well with data and shows the same 
$q_{\rm min}$, while the minimum is shifted to higher $q$ values in {\sc 
Herwig}.

The absence of the oscillations disagrees with earlier results from 
different reactions \cite{hqexp} 
 and in particular with those from e$^+$e$^-$ 
collisions  
at the same energy (SLD \cite{sld}) 
(see also \cite{multrev}).
Since it has been suggested \cite{trunc} 
that the oscillatory behaviour of the $H_q$ could be affected 
by the multiplicity distribution truncation\footnote{This 
already 
occurs naturally as 
a 
consequence of the limited size of the number of events and the 
multiplicity per event.} (SLD), the 
multiplicities 
with relative error greater than 50\% in the multiplicity distribution  
were rejected ($\sim$0.005\% of events).  

\begin{figure}[t]
\resizebox{0.47\textwidth}{0.62\hsize}{%
  \includegraphics{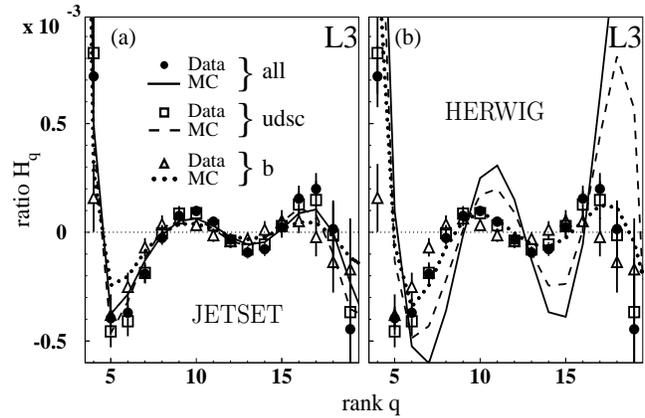}
}
\caption{The $H_q$ moments of the {\it truncated} charged-particle 
multiplicity 
spectra for all, udsc-, and b-quark events of L3 data compared to the 
predictions of 
(a) {\sc Jetset}  and (b) 
{\sc Herwig} 
Monte Carlo 
models, as reported in \cite{l3h}.
}%
\label{fig:l3hf}      
\end{figure}

The $H_q$ ratios for the truncated multiplicity are measured for all, 
udcs-, and b-quarks as shown 
in Fig. \ref{fig:l3hf}.  Despite the moments have a first minimum at 
$q=5$, they {\it do} exhibit quasi-oscillatory behaviour for higher 
ranks. 
No sensitive
differences are visible  among all three samples. 
The  oscillations  are  
well 
reproduced by {\sc Jetset}, while the minima are shifted to higher $q$s 
and are larger in {\sc Herwig}.

L3 concludes that the measurements 
qualitatively
agree with MLLA and NNLO, i.e. show  a negative minimum at 
$q=5$, but 
do not confirm the oscillatory behaviour  predicted by high-order NLO.
The measurements are well described by {\sc 
Jetset}.

Recently, in \cite{l3t} it has been shown that from theoretical 
point of view the {\it un}truncated moments have to be a subject under 
investigation and not the truncated ones.
The very small values of the oscillation amplitudes of 
untruncated 
$H_q$s are shown to follow well the MLLA predictions.

\section{Inclusive $f_1$(1285) and $f_1$(1420) production } 
DELPHI reports on the  hadron spectroscopy measurement of the 
inclusive 
production of two 
$(K{\overline K}\pi)^0$ states in the mass range 1.2--1.6 GeV/$c^2$ in 
hadronic Z$^0$ decays. The measurements are based on the neutral 
$K{\overline K}\pi$ channel in the reaction Z$^0 \to K_SK^{\pm}\pi^{\mp} 
+X^0$, where the two 3-body states in the channel are 
ascribed to  $f_1$(1285) 
and $f_1$(1420) mesons. 

The $f_1$(1285)   
and $f_1$(1420) are the mesons belonging to the $P$-wave hadron multiplets 
$^3P_1$ which (along with the $^1P_1$) are very little studied in contrast 
to 
the well established $S$-wave  ($\pi$, $\rho$) and other $P$-wave,  
$^3P_2$ and  $^3P_0$ (e.g. $f_2$(1270), $f_0$(980)), mesons \cite{pdg}. 
Given the complexity of quark content and possible states to exist in the 
$(K{\overline K}\pi)^0$ systems, the study requires 
high accuracy in the 
selection and analysis procedures available at LEP. To note is that this 
is the first study of the inclusive production of two $J^{PC}=1^{++}$ 
mesons.

A data sample of 3.4M events was processed. After the hadronic
event selection,  specific requirements on the tracks were imposed to
extract the resonances for the $K_SK^{\pm}\pi^{\mp}$ system.
The only events containing 
at least one $K_SK^+\pi^-$ or $K_SK^-\pi^+$ combination are used in 
the analysis, corresponding to a sample of about half a million 
events.
In the study the two methods,  $K_SK^{\pm}\pi^{\mp}$ mass spectra and the 
partial wave analysis (PWA), are applied.
To maximize both $f_1$ mesons signals over background, the data were
estimated
using Monte Carlo events with a mass cut $M(K_SK^{\pm})\leq 
1.4$~GeV/$c^2$.

\begin{figure}[t]
\begin{center}
\resizebox{0.47\textwidth}{.765\hsize}{%
  \includegraphics{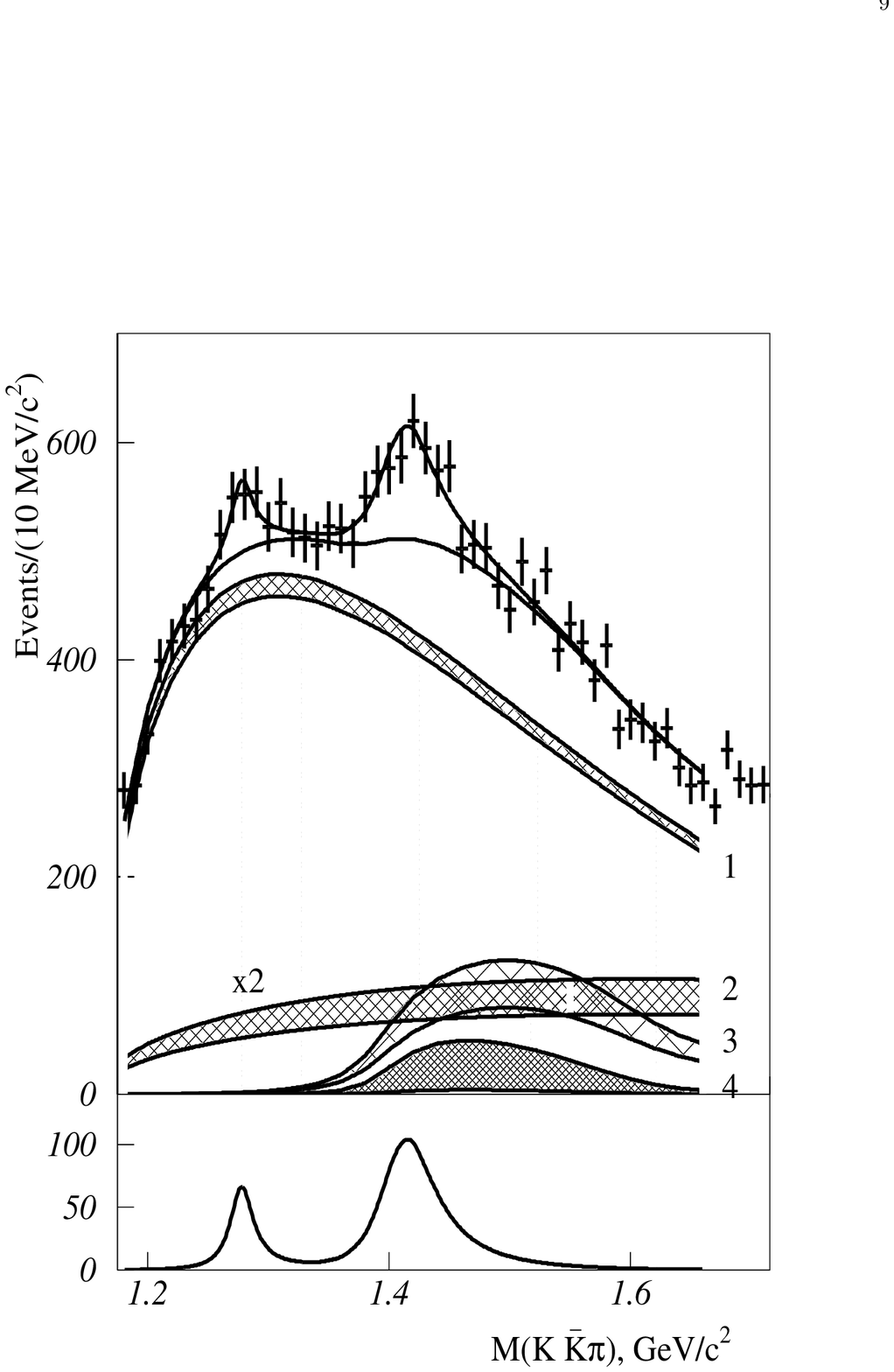}
}
\end{center}
\caption{
$K_SK^{\pm}\pi^{\mp}$ invariant mass distribution measured by 
DELPHI \cite{delf}  with a breakdown into the 
partial waves for the signals (lower histogram) and  the 
background (one error band).  The signals consist of 
$J^{PC}=
1^{++}a_0(980)\pi$ 
(first peak) and $1^{++}K^*(892){\overline K}$ (second peak). The 
background includes: (1) isotropic phase space distribution, and 
(shown magnified by a factor of two) 
the 
partial waves 
of (2) $0^{-+}a_0(980)\pi$, (3) 
$1^{++}K^*(892){\overline 
K}$, and (4) $1^{+-}K^*(892){\overline K}$. 
} 
\label{fig:dpwa}
\end{figure}

The $(K{\overline K}\pi)^0$ mass spectra are fitted in the region 
1.19 to 1.7 GeV/$c^2$  with a 
two $S$-wave 
Breit-Wigner forms and a specific background 
function. From the fits the masses and widths for the two $f_1$ mesons are 
estimated to be, respectively, 
$1274\pm 6$ and $29\pm12$ MeV/$c^2$ for $f_1$(1285)
and 
$1426\pm 6$ and $51\pm14$ MeV/$c^2$ for 
$f_1$(1420), where the errors are the total, statistical and 
systematic, ones.
The corresponding  efficiencies for the two $f_1$s 
are (in \%): $(63\pm 3)$$\times$$10^{-3}$  and $(45\pm 
2)$$\times$$10^{-2}$  (Monte Carlo estimate). 

To get more information on the spin content of the two signals a 
mass-dependent 3-body PWA \cite{pdg} is applied to the 
$K_SK^{\pm}\pi^{\mp}$ system. The Dalitz plots with integrating over 
three Euler angles are fitted providing the 
contribution of the various $J^{PC}$ waves as a function of 
$M(K_SK^{\pm})$. The comparison between fits uses their maximum 
likelihood 
values and their description of the $(K{\overline K}\pi)$, 
$(K\pi)$ and 
$(K{\overline K})$ mass distributions.
The best fit with 
the estimated possible background contributions  are
shown in Fig. \ref{fig:dpwa} and the values obtained are consistent with 
the 
values obtained from the mass spectra study. The (major) systematic 
uncertainties   come from the background description and the PWA fit 
conditions. To estimate them, different fits have been carried out.
A PWA of the $(K{\overline K}\pi)^0$ system shows that the first peak is 
consistent with the $I^G(J^{PC})=0^+(1^{++})a_0(980)\pi$ or 
$0^+(0^{-+})a_0(980)\pi$ waves and the second with the 
$I^G(J^{PC})=0^+(1^{++})K^*(892){\overline K}+ {\rm c.c.}$

The analysis of the measured hadronic production rates per hadronic Z 
decay, $0.165\pm0.051$ and $0.056\pm0.012$, respectively,   for the lower 
and for the 
higher mass signals obtained, suggests that their quantum numbers are very 
probably
$I^G(J^{PC})=0^+(1^{++})$. The comparison of the present 
measurements to the LEP averaged total production rates per spin state and 
isospin for  scalar, vector and tensor mesons as a function of mass 
suggests, in its turn,  
the two mesons quark constituents to be mainly 
u${\overline {\rm u}}$
and   
d${\overline {\rm d}}$.
All this confirms that the measured states are very likely 
$f_1$(1285) and $f_1$(1420) mesons. 

\begin{acknowledgement}
I am thankful to the EPS-HEP2003 Organising Committee, to convenors of the 
Hadronic Physics
Session,
and to my colleagues at CERN and particularly  in OPAL  for giving me
the opportunity to give this talk and
for their kind assistance and support.
\end{acknowledgement}


\begin{thebibliography}{}
\bibitem{opalbl}
OPAL Collab., G. Abbiendi, et al., Phys. Lett. B {\bf 550}, 33 
(2002), 
EPS-HEP2003 abs. 763.
\bibitem{qcdrev}
V.A. Khoze, W. Ochs,  Int. J. Mod. Phys. A {\bf 12}, 2949 (1997).
\bibitem{qcdh1} B.A. Schumm, et al., Phys. Rev. Lett. {\bf 69}, 3025 
(1992).
\bibitem{qcdh2} V.A. Petrov, A.V. Kisselev, Z. Phys. C {\bf 66}, 453 
(1995); Nucl. Phys. B (Proc. Suppl.) {\bf 39B}, 364 (1995).
\bibitem{qcdh3} J. 
Dias de Deus, Phys. Lett. B {\bf 355}, 539 (1995).     
\bibitem{naive} Mark II Collab., P.C. Rowson, et al., Phys. Rev. Lett. 
{\bf 54}, 2580 (1985); A.V. Kisselev, V.A. Petrov, O.P. Yushchenko, Z. 
Phys. C {\bf 41}, 521 (1988).
 \bibitem{delbl}
DELPHI Collab., P. Abreu, et al., Phys. Lett. B {\bf 479}, 118 
(2000),  
{\bf 492}, 398(E) (2000); DELPHI 2002-052 CONF 586.
\bibitem{l3h}
L3 Collab., L3 Note 2808, 2003, Phys. Lett. B (to appear), EPS-HEP2003 
abs. 190. 
\bibitem{hq1}
I.M. Dremin, Phys. Lett. B {\bf 313}, 209 (1993).
\bibitem{correv}
E.A. De Wolf, W. Kittel, I.M. Dremin, Phys. Rep. {\bf 270}, 1 (1996).
\bibitem{multrev}
I.M. Dremin, J.W. Gary, Phys. Rep. {\bf 349}, 301 (2000).
\bibitem{lphd} Ya.I. Azimov et al., Z. Phys. C {\bf 27}, 65 (1985), {\bf 
31}, 213 (1986). 
\bibitem{hqexp} I.M. Dremin, et al.,  Phys. Lett. B {\bf 336}, 119 (1994).  
\bibitem{sld} SLD Collab., K. Abe, et al., Phys.Lett. B {\bf 371}, 149 
(1996).  
\bibitem{trunc} A. Giovannini, S. Lupia, R. Ugoccioni, Phys. Lett. B {\bf 
342}, 387 (1995).
\bibitem{l3t}
M.A. Buican, C. F\" orster, W. Ochs, Eur. Phys. J. C {\bf 
31}, 57 (2003), hep-ph/0307234.
\bibitem{delf}
DELPHI Collab., J. Abdallah, et al., Phys. Lett. B {\bf 569}, 129 
(2003), 
EPS-HEP2003 abs. 323.  
\bibitem{pdg} Particle Data Group, K. Hagiwara, et al., Phys. Rev. D 
{\bf 66}, 010001 (2002). 
\end{thebibliography}
\end{document}